\documentclass [12pt,twoside]{article}           

\usepackage{epsfig,times,lscape}
\usepackage[usenames]{color}

    \ifnum\count102=1

    \fi

    \ifnum\count102=2
\fi

\newcommand{\nc}{\newcommand}

     \ifnum\count101=1
\nc{\qI}[1]{\section{{#1}}}
\nc{\qA}[1]{\subsection{{#1}}}
\nc{\qun}[1]{\subsubsection{{#1}}}
\nc{\qa}[1]{\paragraph{{#1}}}

\def\qpar{\vskip 2mm plus 0.2mm minus 0.2mm}
\def\qL{\hfill \break}
     \fi 

      \ifnum\count101=2
 \nc{\qI}[1]{\parindent=0mm \vskip 8mm 
{\centerline{\LARGE \color{red}#1}}\vskip 3mm}
\nc{\qA}[1]{\vskip 2.5mm \noindent {{\bf        #1}} \vskip 1mm
\parindent=0mm}
 \nc{\qun}[1]{\vskip 1mm \noindent {\sl #1 }\quad }

\def\qL{\hfill \break}
\def\qpar{\vskip 2mm plus 0.2mm minus 0.2mm}

      \fi

\def\qth{\vrule height 12pt depth 0pt width 0pt}
\def\qtb{\vrule height 0pt depth 5pt width 0pt}

\nc{\qfoot}[1]{\footnote{{#1}}}

      \ifnum\count101=1
\def\qbu{\hfill \par \hskip 6mm $ \bullet $ \hskip 2mm}
\def\qee#1{\hfill \par \hskip 6mm (#1) \hskip 2 mm}
      \fi
      \ifnum\count101=2
\def\qbu{\hfill \par \hskip 4mm $ \bullet $ \hskip 2mm}
\def\qee#1{\hfill \par \hskip 4mm (#1) \hskip 2 mm}
      \fi

\def\qparr{ \vskip 1.0mm plus 0.2mm minus 0.2mm \hangindent=10mm
\hangafter=1}

     \ifnum\count101=1 
 
     \fi
     \ifnum\count101=2

  \def\qcitb#1{\noindent \hbox to 102mm{\hfill \small #1} \vskip 1mm}
      \fi

 \def\qpages#1{\count102=0{\loop\advance\count102 by 1
 \null \vfill\eject \ifnum\count102<#1 \repeat}}

\def\qth{\vrule height 12pt depth 0pt width 0pt}
\def\qtb{\vrule height 0pt depth 5pt width 0pt}

\def\qv{\vskip 0.1mm plus 0.05mm minus 0.05mm}

\def\qhw{\hskip 1.5mm}
\def\qleg#1#2#3{\noindent {\bf \small #1\qhw}{\small #2\qhw}{\it \small #3}\qv }
\begin{document}
\thispagestyle{empty}



\markboth{{\sl \hfill  \hfill \protect\phantom{3}}}
        {{\protect\phantom{3}\sl \hfill  \hfill}}

\color{yellow} 
\hrule height 20mm depth 10mm width 170mm 
\color{black}
\vskip -2.0cm 
\centerline{\bf \LARGE Clustering experiments}
\vskip 1.8cm
\centerline{\bf \large 
Zhengwei Wang$ ^1 $,
Lei Wang$ ^2 $,
Ken Tan$ ^{1,3} $,
Zengru Di$ ^4 $,
Bertrand M. Roehner$ ^{4,5} $ }

\large
\vskip 1cm

{\bf Abstract}\quad It is well known that bees 
cluster together in cold weather, in the process of
swarming (when the ``old'' queen leaves with {\it part} of the
colony) or
absconding (when the queen leaves with {\it all} the colony)
and in defense against intruders
such as wasps or hornets.\qL
In this paper we describe a fairly different
clustering process which occurs at any temperature
and independently of any special stimulus or circumstance.
As a matter of fact, this process is about four times
faster at 28 degree Celsius than at 15 degrees.\qL
Because of its simplicity and low level of ``noise''
we think that 
this phenomenon can provide a means for exploring
the strength of inter-individual
attraction between bees or other living organisms. \qL
For instance, and at first sight fairly surprisingly, 
our observations showed that
this attraction does also exist between bees belonging to
{\it different} colonies.  \qL
As this study is aimed at providing
a comparative perspective, we also describe a similar clustering
experiment for red fire ants.

\vskip 8mm
\centerline{15 December 2011}
\vskip 2mm
\centerline{\it Preliminary version, comments are welcome}

\vskip 6mm
{\normalsize Key-words: clustering, aggregation, 
condensation, attraction, 
coupling strength, interaction, temperature, critical density,
bees, colony.}
\vskip 10mm

{\normalsize 
1: Eastern Bee Institute, Yunnan Agricultural University, Kunming,
China. \qL
2: Laboratory of Insect Ecology, 
Red Imported Fire Ants Research Center, South China Agricultural
University, Guangzhou, China.\qL
3: Corresponding author, email address: eastbee@public.km.yn.cn.\qL
4: Department of Systems Science, Beijing Normal University, Beijing,
China. \qL
5: Corresponding author, email address: roehner@lpthe.jussieu.fr.\qL
On leave of absence from the ``Institute for Theoretical and
High Energy Physics'' of University Pierre and Marie Curie, Paris, France.
}

\vfill \eject

\qI{Introduction}

The paper describes a clustering process through which spatially
dispersed bees aggregate into a single cluster. 
This is similar to the condensation process through
which molecules of vapor come closer together
to form droplets of liquid water.
We think that such clustering processes 
may give information about
inter-individual coupling strength in a population, just like data
about phase transition provide information on
inter-molecular interactions.
\qpar

After the present introduction there 
will be three parts in this paper. (i) First we describe
a typical clustering experiment so that such experiments may
be repeated (and possibly expanded) by other researchers.
(ii) Secondly, we give some preliminary results.
(iii) Finally, we list a number of questions which we plan to explore
in forthcoming experiments.
\qpar

But before we start let us answer a fairly natural question, namely
what is the
rationale for {\it measuring} coupling strengths in animal or 
human populations, which indeed is our final objective.
It is fairly obvious that the bond
between children and their parents is stronger than the link
between colleagues working in the same company but is it
two times or ten times stronger? We do not know because, so far,
there is no experimental procedure for probing the strength of
such links%
\qfoot{Let us emphasize that the time scale does matter here.
It would be fairly meaningless to try to answer such a question
on a daily basis because there would be
big (and more or less random) fluctuations. On the contrary, 
on a time scale of several decades, the connections inside a family
are likely to be maintained with more stability than those 
between colleagues. In addition to time averaging there
should also be an average over a sufficiently large 
population. For measurements of inter-molecular interactions 
these averaging processes are done, so to say, 
automatically because any
sample will include of the order of $ 10^{23} $ molecules
and the time-scale of the fluctuations in their interactions
is of the order of $ 10^{-10} $ second.}%
.
This simple example emphasizes the fact that
in contrast to physicists, biologists or sociologists have
no quantitative knowledge whatsoever about interaction strength.
\qpar

But why is it important to explore the strength of bonds? From
physical chemistry we know that 
intermolecular coupling strength
is the determining factor of most physical properties of any compound.
For instance, 
as is well known, water molecules will form vapor, drops of liquid
or blocks of ice depending on the coupling strength between them.
In these three states
the mechanisms of interaction between molecules are basically
the same, only their average distances
and their spatial organizations differ.
The same conclusion holds for many other 
important physical properties, e.g.
density, boiling temperature, latent heat, equilibrium vapor pressure,
speed of sound. In short, coupling strength is key to
characterizing the main properties of a system.
\qpar

Populations of insects (or of small animals such as fruit flies or
small fishes) provide a convenient testing field for experimental
methods and devices designed for measuring coupling strength.
Intuitively, one would expect social insects to have stronger
bonds than insects which do not live in colonies. 
As it is probably easier to measure strong bonds rather than
weak bonds, it may be a good strategy to begin by studying
insects which have strong social organizations such as
honey bees or some species of ants.
\qpar

That is why our attention was particularly attracted by a paper published
in 1950 by a French naturalist (Lecomte 1950) which describes a
clustering process in a (sufficiently large) population of bees.
The first step in our investigation was to repeat this experiment.
This is described in the next section.

\qI{Description of a clustering experiment}
The experiment described below
was performed on 16 November 2011 at the ``Eastern Bee Institute''
of Yunnan Agricultural University, Kunming, China.
\qee{1} At around 1pm during a sunny and fairly windy afternoon
some $ n=120 $ bees%
\qfoot{{\it Apis cerana} also called Eastern honey bees. 
The {\it cerana} bees
have an average body length of about 12mm which means that they are
slightly shorter than western honey bees 
({\it Apis mellifera}) which on average have a length of 14mm.}
were  transferred by aspiration
from the frames of a beehive into a plastic bottle.
\qee{2} Immediately after being collected the bees were put
to sleep by flooding the bottle with carbon dioxide during about
2 minutes.
\qee{3} Then the bees were spread on the bottom of a box of size
(a=22cm, b=29cm) as shown in Fig. 1. In this experiment we did not try
to distribute the bees uniformly on the vertex of a lattice because
we wished to follow the same procedure as described by Lecomte 
(1950). \qL
This repartition corresponds to a density 
$ d=N/S=N/ab=18.7 $ bees per square decimeter. 
A more suggestive parameter is the average spacing, $ e $,
between closest neighbors.
For a density $ d $ each bee occupies an
area $ s= S/N =1/d=5.3 $ square centimeter.
In other words, the average distance between 
neighboring bees is $ e=\sqrt{s}=1/\sqrt{d}=2.3 $cm.
%
\count101=1
\ifnum\count101=1
\begin{figure}[htb]
\vskip -7mm
 \centerline{\psfig{width=12cm,angle=-90,figure=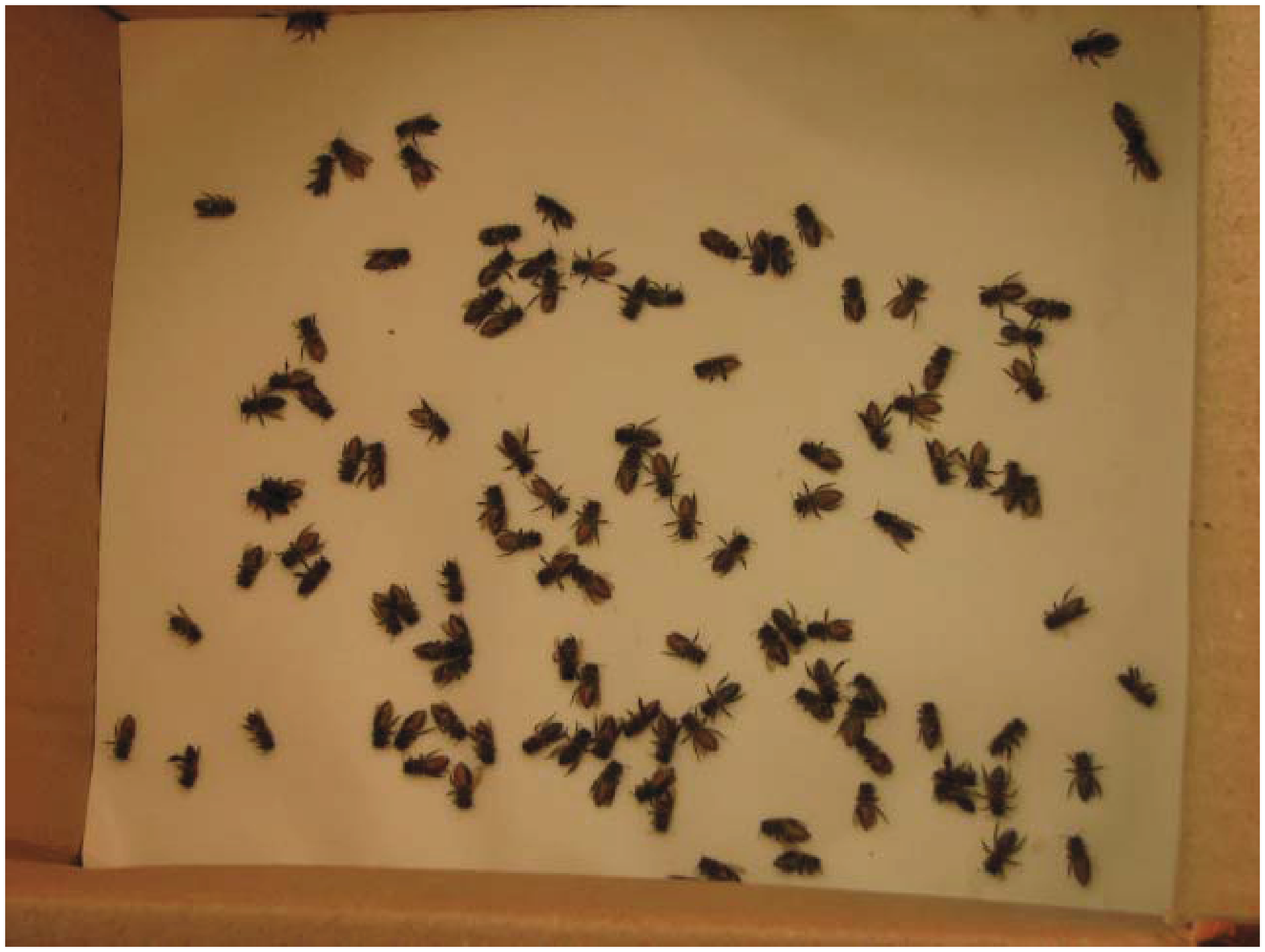}}
\vskip -10mm
\qleg{Fig. 1: Clustering experiment: initial distribution of bees.} 
{Some 120 {\it cerana} bees have been put to sleep with carbon
dioxide and spread on the bottom surface of a container.
The average spacing between nearest neighbors is 2.3cm.
The temperature in the container is around 28 degree Celsius.}
{}
\end{figure}
\fi
%
\qee{4} The box was put outside in sunshine 
but closed with a board of wood
so that the bees were in the dark.
The experiment started at this point. Time was 1:40pm.
Every 10 minutes the box was opened for a short moment (around 20 s)
for a picture to be taken.
The temperature inside of the box was monitored thanks to a
digital thermocouple thermometer. 
On average it was around 28 degree 
Celsius with an upward trend due to the sunshine.
\qee{5} At 1:50pm there were about 8 small clusters comprising 
7 to 15 bees; some 25 ``isolated'' bees were not yet
part of any cluster.
\qee{6} At 2pm there were 4
clusters as shown in Fig. 2. Each of the two large clusters
 had some 35 bees whereas the two smaller clusters had 15 bees.
Clearly, the reduction by a factor two in the number of clusters was
achieved through the coalescence of neighboring clusters.
%
\ifnum\count101=1
\begin{figure}[htb]
\vskip -7mm
 \centerline{\psfig{width=12cm,angle=-90,figure=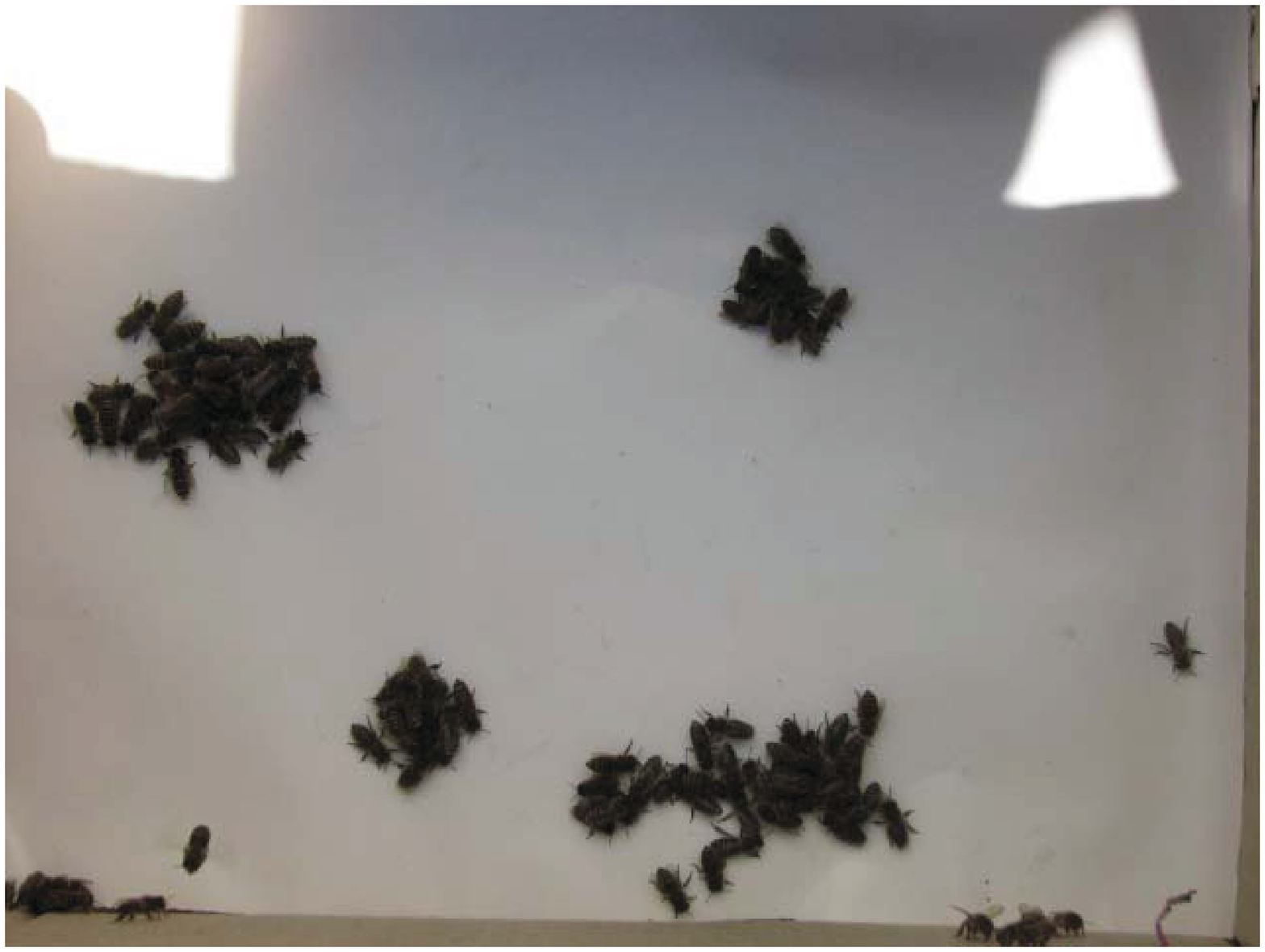}}
\vskip -10mm
\qleg{Fig. 2: Clustering experiment after 20 minutes.} 
{In the process which is under way the bees had first 
aggregated into many small groups and then these groups
have been moving toward one another to form larger
clusters until forming one single cluster 
at the end of the experiment (as shown below in Fig. 3).}
{}
\end{figure}
\fi
%
\qee{7} At 2:10pm, that is to say, half an hour after the
experiment started, 
only two clusters remained: a big one
which was moving (globally) on one of the side walls of the box and a
much smaller one which remained at the bottom.
\qee{8} When the box was opened again at 2:20 a large number of
the bees flew out of the box. In this respect it should be noted 
that even shortly after the start of the experiment a few bees
managed to fly away%
\qfoot{For a physicist this does not come as a surprise.
Of course, if one wishes,
one can try to imagine some special reasons.
For instance, it may be that the repartition of carbon dioxide
in the bottle
was not uniform with the result that some bees 
did almost not sleep. But one can also take a more
global perspective by observing that some molecules are able to escape
even from a block of ice as shown by the fact that for ice at -10
degree Celsius the equilibrium pressure
is 2.6 mbar. This means that $ 2.6/1000=0.26\% $ of
the molecules have been able to escape. For the sake of comparison
at +10 degree the vapor pressure is 12 mbar, that is to say 
only 5 times higher than over ice.}%
.

\qI{Results}

The present paper summarizes only the first phase of our investigation,
but, however limited, the experiments that we have conducted so far
lead us to three useful conclusions.
\qee{1} The aggregation effect can be observed repeatedly with only
small variability.
\qee{2} This aggregation effect has nothing to do with the
cluster formation that occurs in cold weather.
\qee{3} The present aggregation phenomena gives a key for 
probing the attraction strength between bees or other living
organisms.
\qpar
The first point is perhaps of particular importance because
biological experiments are often plagued by a high level of noise.

 \qA{Speed of clustering increases with temperature}
When we discussed the experiment described in the paper by Lecomte
(1950) with other researchers several of them suggested that
the clustering effect may be a reaction to low temperature.
This was indeed a natural comment for it is well known that bees
cluster to increase the temperature in the middle of the cluster.
In such cases they do not only form a cluster but they also activate
their muscles (and especially the strong muscles of their wings)
so as to generate heat.
Such a clustering may have different purposes. 
\qbu One purpose is to keep the temperature of the brood at 
the required level of 35 degree
Celsius in spite of a temperature outside of
the beehive which may be much lower.
\qbu A different clustering purpose is 
a defense tactic against hornets or wasps.
When a hornet or a wasp tries to break into a beehive
of {\it cerana} or {\it mellifera}
a ball of bees surrounds the invader. The bees
vibrate their flight muscles until
the temperature inside of the ball 
is raised to 47 degree Celsius. 
Together with an elevated concentration of carbon dioxide this
temperture kills the intruder but does not harm the bees
because their lethal temperature is higher. It can
be noted that fever in humans is a similar tactic against
viruses%
\qfoot{The account of this defense tactic given in the 
Wikipedia article entitled {\it Apis cerana} (English
version) is not really
reliable in several respects. First, he does not cite
the original paper by M. Ono, I. Okada and M. Sasaki (1987).
Secondly, it says that this defense tactic is specific to the
{\it cerana} bee while in fact, as shown by K. Tan et al. (2005),
it is also used (albeit less effectively) by the {\it mellifera}.
Thirdly, it does not mention the role of oxygen deprivation
which was shown to be an important factor by
M. Sugahara and F. Sakamoto (2009);
moreover 
it gives the lethal temperature of {\it cerana} bees
as being 49 degrees while it seems closer to 51 degrees
according to the same authors.
Finally, it does not mention that
the same kind of defense is used not only against giant hornets but 
also against wasps (see K. Tan et al. 2005, 2010). \qL
Incidentally,
what is not completely clear in the defense tactic against the giant
Japanese hornet is
why the hornet does not simply kill
the bees which surround it in order to break this 
deadly surrounding.
Indeed, the Wikipedia article 
about this giant hornet says that it can kill as many as 40 bees
per minute. Thus, it should be able to ``drill'' an
escape tunnel through the ball in less than one minute,
that is to say much less than the
10 mn that it takes for it to be killed by the ball.}%
.
\qpar

In these two cases, the purpose of the cluster is to provide
thermal isolation and to concentrate the production of heat.
In order to test whether or not the clustering described by Lecomte
is of the same type we performed the same experiment at two different
temperatures. 
\qbu A first experiment was conducted at 15 degree Celsius.
Clustering occurred but took a long time, about 2 hours.
At such a low temperature one might at first suspect that the clustering
was indeed an attempt to keep heat inside of the cluster.
However, the bees did not 
vibrate their flight muscles which means that they did not
try do raise the temperature of the cluster.
\qbu A second experiment was done at 28 degree Celsius. As this
temperature is already higher than the normal temperature
of the bees in the beehive (that is to say the temperature
between the frames) there is certainly no need to cluster
to prevent a loss of heat. Nonetheless, clustering occurred
and was in fact about 4 times faster than at 15 degree. 

\qA{Clustering in a mixed population}
In this experiment, some hundred {\it cerana} bees were collected
from a beehive $ A $, and another hundred from a beehive $ B $ located
some 15 meters from $ A $. All the $ A $ bees were marked with a white
dot. Then,
the two populations were put to sleep
and dispersed at the bottom of a box 
as explained in the previous section. The box had a area
which was twice the area of the previous box so as to keep the
same population density. The question was whether the $ A $ and $ B $
bees would form separate clusters or whether they would form 
mixed clusters.
\qpar
It is a common belief that each colony has its own odor and that
if a foraging bee from $ A $ tries to enter the beehive $ B $ it
is identified at the entrance, prevented from entering
and possibly even killed. One of the US researchers with whom we were
in contact before doing this experiment
emphasized the crucial role of odors in the communication 
between bees.
\qpar

Yet, observation instead showed that the bees form a 
mixed cluster (see Fig. 3).
%
\ifnum\count101=1
\begin{figure}[htb]
 \centerline{\psfig{width=7cm,angle=90,figure=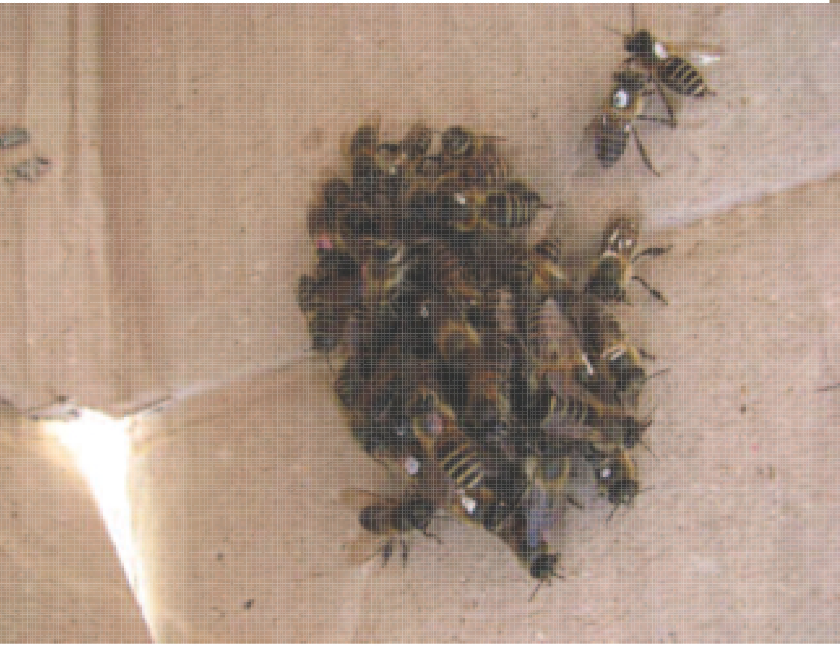}}
\vskip 2mm
\qleg{Fig. 3: Mixed cluster of bees.} 
{The bees from colony $ A $ are marked with a white dot whereas
those of colony $ B $ have no dot. The picture
shows that the cluster comprises
bees from the two colonies.}
{}
\end{figure}
\fi
%
This result is not completely unexpected because beekeepers
know that it is possible to put frames from different colonies 
into the same beehive. Fights may be prevented 
by spraying all the bees with flavored water.

\qA{Possible role of vibrations}
In his paper of 1950, J. Lecomte describes a second experiment
which is the following. On the bottom of the box where the
bees were scattered, he disposed a cell containing at least 100 bees.
Then he found that, once formed, the whole cluster would slowly move 
toward this cell (at a speed of about 6\ cm/hour)
until covering it completely.  
\qpar
In a connected paper Lecomte (1949) made an observation which may 
give a clue as to the interaction mechanism. He observed that if 
a wad of cotton is interposed between the 
cell containing the bees and the bottom of the box the cluster
does no longer converge toward the cell. Lecomte attributes this
result to the fact that the vibration of a frequency around 30Hz that
the bees are known to produce is damped by the cotton.
In his article of 1950 he says that he tried to simulate
this effect by generating a vibration of same frequency with an
electro-mechanical device but it failed to attract the cluster 
of bees.

\qA{Role of the number of bees}
In a subsequent article published in 1956, J. Lecomte investigated
the role of the 
%
\begin{table}[htb]

\centerline{\bf Table 1\quad Influence of the number 
(and density) of bees on cluster formation}

\vskip 2mm

\hrule
\vskip 0.5mm
\hrule

\vskip 2mm
$$ \matrix{
\qtb
 \hbox{Number of bees} \hfill & 5 & 10 & 15 & 25 & 50 & 75 \cr
\noalign{\hrule}
\qth \qtb
\hbox{Frequency of formation of a cluster}\hfill &
                     30\% & 40\% & 6\% & 6\% & 80\% & 100\% \cr
\noalign{\hrule}
} $$
\vskip 2mm
{
Notes: The experiments were performed at 25 degree Celsius.
For each total number of bees the experiment was repeated 18 times.
The same box was used in all experiments which means that
the density of the bees per square centimeter decreased along with the
number of bees. A cluster was defined as an aggregate
containing at least 80\% of the 
total number of bees. \qL
The results show a fairly sharp
transition between 25 and 50 bees. \qL
There
is no clear explanation for the fact that the probability
increases again for 5 and 10 bees; of course, for such small numbers
the variability may be large which means that 
in addition to the average one would also
need to know the standard deviation.
\qL
Source: Lecomte (1956).}
\vskip 2mm
\vfill

\hrule
\vskip 0.5mm
\hrule
\end{table}
%
number of bees in a more quantitative way than he
had done in his papers of 1949 and 1950. He carried out
6 series of experiments with 5, 10, 15, 25, 50, 75 bees respectively.
In each series the experiment was repeated 18 times which resulted
in a total of $ 6\times 18=108 $ experiments.
He considered that a valid cluster had to include 
at least 80\% of the bees.
With this definition he obtained the results given in Table 1.
\qpar

What is not really satisfactory in this experiment is the fact
that the number and the density of bees change together.
In order to determine which of these two variables is the
determining factor, one needs to perform two series of experiments.
\qbu In one series the spacing between the bees would be kept
constant but their number would be decreased 
until the clustering process disappears.
\qbu In a second series of experiment the number would be kept 
constant but the spacing between the bees
would be progressively increased until
the clustering process disappears.
\qpar

This requires a fairly broad study that
we are planning to conduct in the near future.

\qI{Clustering experiment for ants}

In order to emphasize our commitment to comparative analysis
we describe in this section a clustering experiment involving
ants. It was carried out 
with red imported fire ants ({\it Solenopsis invicta} Buren)
in the summer of 2011 at the 
``Laboratory of Insect Ecology'' of the South China Agricultural
University.

\qA{Description of the experiment}

The experiment involved two steps.
%
\begin{figure}[htb]
 \centerline{\psfig{width=12cm,figure=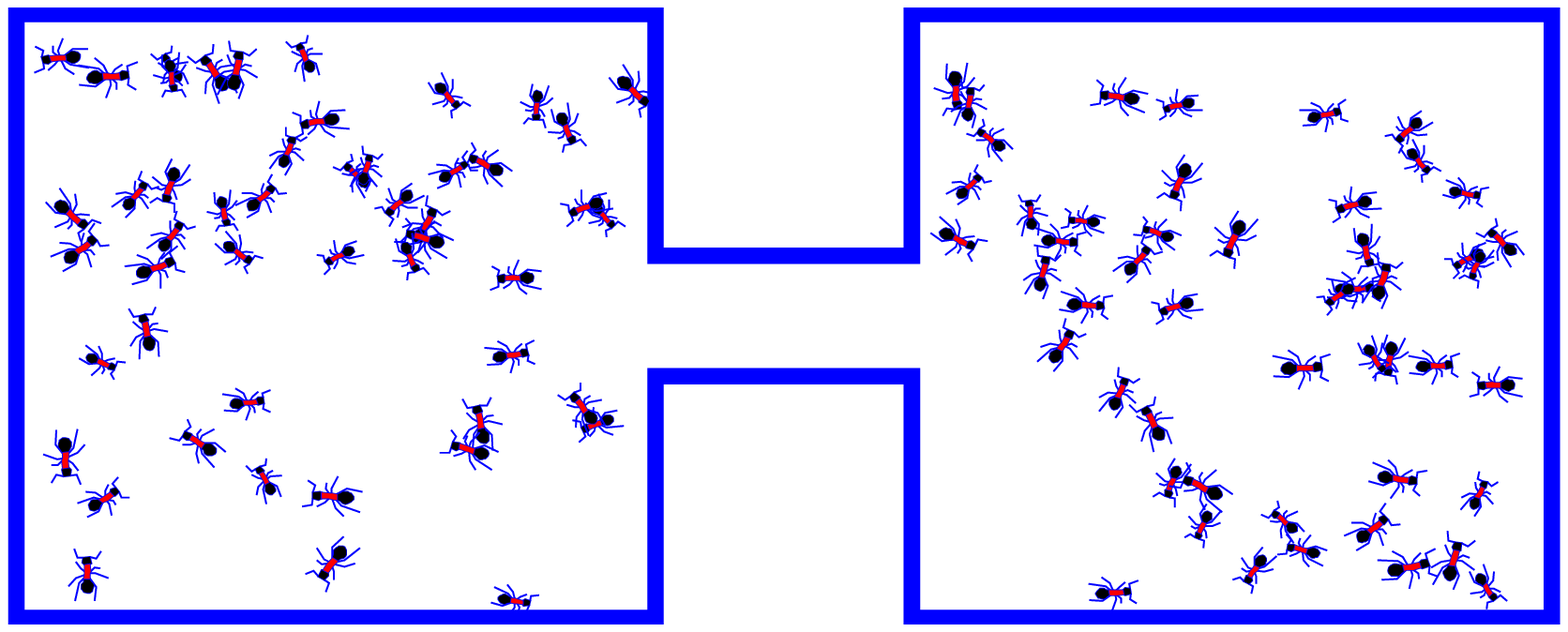}}
\vskip -10mm
 \centerline{\psfig{width=12cm,figure=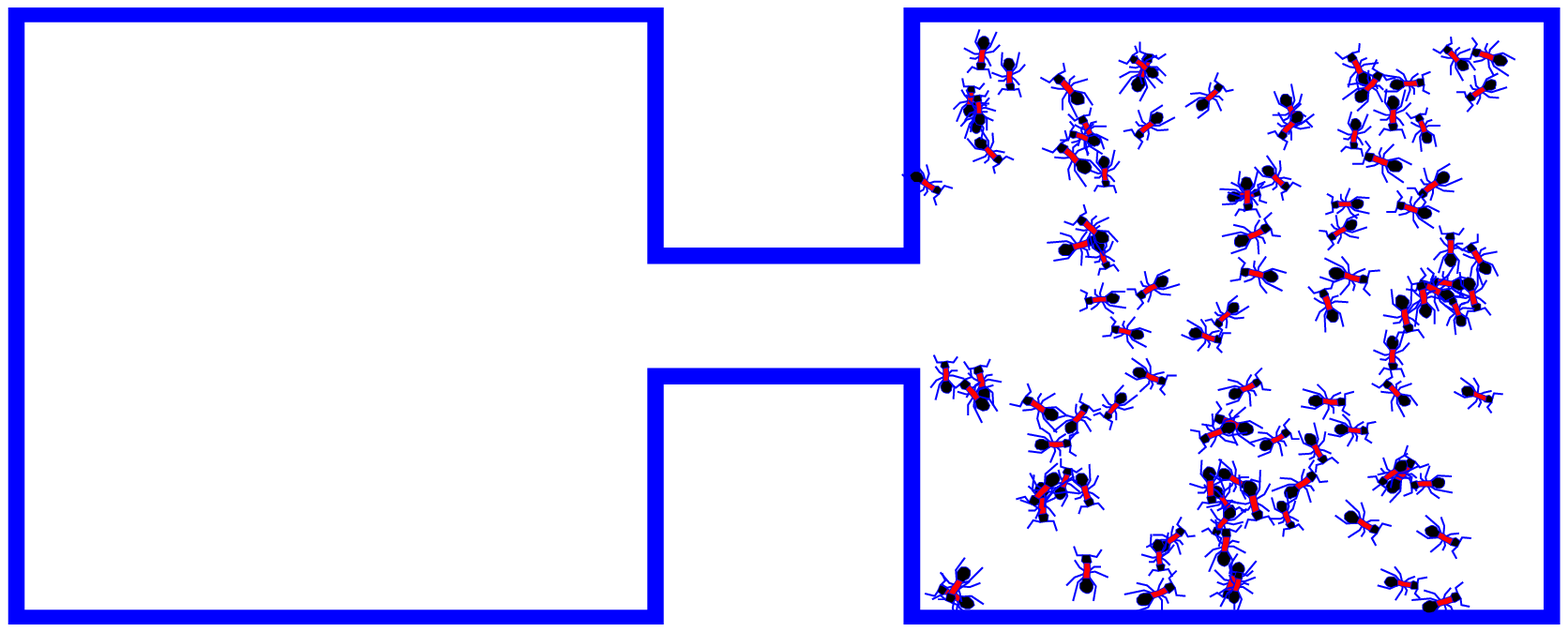}}
\vskip -2mm
\qleg{Fig. 4: Clustering experiment for ants.} 
{The diagram at the top shows the initial state while the
diagram below shows the state after a few hours when all ants
have gathered on the same side.
The ants came from the same colony. They were initially put
to sleep through carbon dioxide.
For the clarity of
the diagram only 50 ants are shown on each side 
in the initial situation.
In the experiment
the initial numbers $ N $ of ants on each side were 100, 200, 300, 500,
750, 1000. \qL
If $ T $ denotes the delay after which
the ants are all in the same box one observes that
the larger $ N $, the longer $ T $; the relation
between the two variables was found to be ($ N $ in hundreds, $ T $
in hours): 
$$ T=aN+b,\quad a=0.5\pm 0.3,\ b=2\pm 1 $$
with a coefficient of linear correlation of $ 0.84 $. }
{}
\end{figure}
%
%
\qbu First a number $ N $ of ants were put in two
flat boxes connected by a glass tube (Fig. 4). The boxes had an area of
about 60 square centimeters. The connecting glass tube had a length of
3 centimeters and a diameter of one centimeter. The ants were initially
put to sleep by using carbon dioxide. Thus, the ants were in
a similar condition as the bees in the previous experiments.
\qbu After a time of the order of several hours the ants gathered
together on the same side. It can be observed that although
the longest experiments lasted of the order of 12 hours,
the ants received no food nor water during the experiment.\qL
Of course, a few ants remained in the
initial box; for instance the ants which did not awake after
being put to sleep; the criterion which was used to terminate
the experiment was that the number of remaining ants should
be smaller than 10\% of the number initially contained
in the box.

\qA{Results}
 For each $ N $ the experiment was repeated 5 times. The variability
was fairly high; the coefficient of variation, 
$ \sigma/m $, of
the times in the 5 repetitions was on average 65\%.
\qpar

For each of the $ N $ the average of the 5 repetitions gave the
following results ($ N $ in hundreds and $ T $  in hours): 
$$ N,T: (1,2.3)\ (2,2.1)\ (3,5.1)\ (5,2.8)\ (7.5,5.9)\
(10,6.8) $$

The relation between $ N $ and $ T $ was found to be:
 $$ T=aN+b,\quad a=0.5\pm 0.3,\ b=2\pm 1 \quad 
\hbox{for}\ 100\le N\le 1,000 $$

\qA{Comments}
Of course, for the sake of comparative analysis, it would 
be interesting to perform exactly the same experiment as with
the bees, that is to say observing what happens when the
ants wake up after being scattered on the bottom of a box.\
This is an experiment that we intend to do in the near future.
\qpar

We are also planning a number of additional experiments.
\qbu What happens when the number $ N $ becomes much smaller
than 100,
for instance of the order of 20 or 30. For bees, one knows 
that there is no longer any clustering for such small numbers.
Is there also such a critical threshold for ants?
\qbu What happens when ants from a colony $ A $ are introduced
in one box and ants from a colony $ B $ in the other box?
Will they nevertheless gather together in the same box?
\qbu What happens when one uses a longer glass tube, for
instance of a length of 30 centimeters instead of 3\ cm?
Will the ants nevertheless gather on the same side%
\qfoot{Preliminary observation shows that if the glass tube 
is longer, many ants will stay in it which will hinder passage
of the other ants from one side to another.}%
?

\qI{Forthcoming research}

Before turning to the future, we give a short account of the
research conducted in past decades, mostly in the time
interval between 1940 and 1975.

\qA{Past research}

The papers published by J. Lecomte were
not an isolated research but were part
of a broader investigation into the collective behavior of
social insects. 
At that time this research was conducted by a group
of French scientists lead by Prof. R\'emy Chauvin (1913-2009).
Chauvin's interest into collective phenomena  started 
with his PhD thesis (1941) which he devoted to the
phase transition through which solitary locusts come to form
large swarms containing up to several billion locusts%
\qfoot{A comparative description of this phase transition for both
animal and human populations
can be found in Roehner (2005, 674-675).}%
.
In the four decades between 1940 and 1980, as the head of different
laboratories, Chauvin and his students and collaborators investigated
several facets of collective behavior among social insects.\qL
In many respects the case of R\'emy Chauvin was similar to 
that of the American psycho-sociologist Stanley Milgram.
Like Chauvin, Milgram investigated collective behavior
and he too saw his work decried in some circles of
academia%
\qfoot{The article that the English version of Wikipedia devotes
to R\'emy Chauvin does hardly justice to his scientific work
in the sense that it focuses almost completely on Chauvin's
researches about parapsychology and unidentified flying objects
that he conducted in the last part of his life.
Why should such topics not be open to legitimate
scientific investigation?
The attitude reflected in this article
rather seems to be a sad reflection on the level of
conformism and narrow-mindedness in 
some areas of present-day scientific research.
Chauvin was also a vocal critic of
neo-Darwinism because he saw it 
as an insufficient yet unfalsifiable theory. A more
honest account of Chauvin's work can be found in the
French version of the Wikipedia article.}%
.
Unfortunately, both Chauvin and Milgram had few (if any) followers.

\qA{The future}

Future research can be directed into two different directions.
\qbu One is the investigation of the {\it detailed} mechanisms 
that play a role in the clustering phenomenon for bees.
\qbu An alternative direction is to make a {\it comparative} study
of clustering phenomena in various conditions and
for different species.
\qpar

Of course, it is impossible to predict in advance which direction will
prove the most fruitful. However, physicists would make the observation
that in physics it was the second option which proved the most useful.
It is easy to explain why if we consider what would be
the parallel of the first option for physical
systems. 
\qpar

Molecules have many mechanisms of interaction: covalent bonding,
ion-ion bonds, dipole-dipole interactions, interaction through
induced dipoles (the so-called London forces). A real investigation 
of these interactions requires that one study them not only at 
molecular level but also at the level of atoms and electrons.
At this level the explanations must rely on quantum mechanics.
In other words, in this direction one faces very difficult 
and tricky problems. One can even say that the answers will
remain fairly elusive because there will always be a more 
detailed level to be considered.
\qpar

In the same way, let us see what would be the second option
for physical systems. It consists in lumping together all
the different kinds of interactions and estimating
the strength of the global attraction forces for instance 
in terms of the energy (expressed in electron-volts)
that it takes to break such bonds.
In this way, as already mentioned at the beginning of the
paper, one is able to predict many important properties
of a great number of liquids or solids.
\qpar

It may be argued that an argument that holds for physical systems
may not necessarily be true for biological systems.
We will not know until we have tried. In this connection
it can also be observed that the comparative approach that we
advocate here was commonly used (both in biology and in the social
sciences)
in the late nineteenth and early twentieth
centuries. It is probably the high degree of 
scientific specialization
that prevails nowadays that killed comparative research.
\qpar

Just as one example of the kind of questions that one would
like to consider in the future one can give the following illustration.\qL
When one opens a beehive one sees hundreds of bees closely packed 
together on
each frame. For {\it cerana} bees
their spatial density may be of the order of 200 per
square decimeter on each side of a frame. To our best 
knowledge this is a far higher density that can be found in a
colony of ants%
\qfoot{However it can be noted that in laboratory colonies
of red fire ants there is also a high density of ants especially
around the water supply.}%
. 
Therefore one would expect a higher attraction
force between bees than between ants. The challenge is to
design an experimental procedure which gives meaningful
attraction estimates in the two cases. It is known that
for some species of ants there are also clustering phenomena%
\qfoot{Many thanks to Dr. Lei Wang for the information
that he gave us about a clustering experiment with
red fire ants that he has conducted. In this experiment
an equal number of ants is put in two boxes connected by a glass
tube. After a few hours all the ants are concentrated on 
only one side.}%
.
Can they be used to get interaction estimates?
\qpar

Although the analysis of clustering is only one among several possible 
methods for measuring interaction strength%
\qfoot{For indications about some other methods see Roehner 
(2008, chapters 4 and 5).}%
,
it seems to be one of the most promising. We hope that this paper
will bring about more researches in this direction.

\vskip 8mm

{\bf Acknowledgements}\quad
We would like to express our gratitude to the researchers 
who have provided advice, help and encouragement in the early stages
of this project and in particular to 
Qing Yun Diao (Bee Research Institute, Chinese Academy of
Agricultural Sciences, Beijing),
Gen Li (Beijing Normal University),
Yan Li (Institute of Biophysics, Chinese Academy of Science, Beijing),
Qingqing Liu (Institute of Biophysics, Chinese Academy of Science, Beijing),
Gene Robinson (University of Illinois at Urbana-Champaign),
Olav Rueppell (University of North Carolina at Greensboro),
David Tarpy (North Carolina State University),
Yijuan Xu (South China Agricultural University).

\qI{References}

\qparr
Chauvin (R.) 1941:
Contribution à l'\'etude physiologique du criquet 
p\`elerin et du d\'eterminisme des ph\'enom\`enes gr\'egaires.
[A study of the physiology of the desert locust (Schistocerca gregaria)
and of gregarious phenomena.]
Soci\'et\'e Entomologique de France, Paris.

\qparr
Chauvin (R.) 1954: Aspects sociaux des grandes fonctions chez
l'abeille. La th\'eorie du superorganisme.
[A societal perspective of bee behavior.]
Insectes Sociaux 1, 123-129.

\qparr
Chauvin (R.) 1972: Sur le m\'ecanisme de l'effet de groupe
chez les abeilles [Analysis of the group effect in bees.
The expression ``group effect'' refers to the fact that
bees live longer when they are part of a large group 
instead of being kept in isolation.]
Comptes Rendus de l'Acad\'emie des Sciences, CRAS
series D, vol. 275, 2395-2397.

\qparr
Lecomte (J.) 1949: L'inter-attraction chez l'abeille.
Comptes Rendus de l'Acad\'emie des Sciences (CRAS), 229,857-858.
[Attraction phenomena among bees. Paper published in the
Proceedings of the French Academy of Sciences.]

\qparr
Lecomte (J.) 1950: Sur le d\'eterminisme de la formation
de la grappe chez les abeilles [About the clustering process
in bees].
Zeitschrift f\"ur vergleichende Physiologie [Journal
for comparative physiology] 32, 499-506. 

\qparr
Lecomte (J.) 1956: Nouvelles recherches sur l'inter-attraction
chez {\it Apis mellifica}%
\qfoot{{\it Apis mellifica} is the same species as {\it Apis mellifera}.
In Latin the meaning
of ``mellifera'' is ``to bear honey'' whereas the meaning of 
``mellifica'' is ``to make honey'' which is of course more
correct because indeed bees absorb nectar and deliver the honey
only once they come back to the beehive.}%
.
[New results about attraction phenomena among {\it Apis mellifica}
(that is to say western honey bees).] 
Insectes Sociaux 3,1,195-198.

\qparr
Ono (M.), Okada (I.), Sasaki (M.) 1987:
Heat production by balling in the Japanese honeybee, 
{\it Apis cerana japonica}
as a defensive behavior against the hornet 
({\it Vespa mandarinia japonica}). Experientia, 43.

\qparr
Roehner (B.M.) 2005: A bridge between liquids and socio-economic
systems. The key-role of interaction strengths.
Physica A, 348,659-682.

\qparr
Roehner (B.M.) 2008: Interaction maximization as an
evolution principle for social systems.
Lectures given at Beijing Normal University in 
September-December 2008.

\qparr
Sugahara (M.), Sakamoto (F.) 2009:
Heat and carbon dioxide generated by honeybees jointly act 
to kill hornets. 
Naturwissenschaften, 96,9,1133-1136.

\qparr
Tan (K.), Hepburn (H.R.), Radloff (S.E.)
Yu (Y.), Liu (Y.), Zhou (D.), Neumann (P.) 2005:
Heat-balling wasps by honeybees.
Naturwissenschaften, 92,492–495.

\qparr
Tan (K.), Li (H.), Yang (M.X.), 
Hepburn (H.R.), Radloff (S.E.) 2010:
Wasp hawking induces endothermic heat production in guard
bees.
Journal of Insect Science, 10,142 (on line publication).

\end{document}